\documentclass{JINST}
\let\ifpdf\relax
\usepackage{tabularx}
\usepackage{booktabs}
\usepackage{amsmath}

\title{Study of the response and photon-counting resolution of silicon photomultipliers using a generic simulation framework}

\author{Patrick Eckert$^a$\thanks{Corresponding
author.}~, Rainer Stamen$^a$ and Hans-Christian Schultz-Coulon$^a$\\
\llap{$^a$}Kirchhoff-Institut f\"ur Physik, Universit\"at Heidelberg,\\
  Im Neuenheimer Feld 227, 69120 Heidelberg, Germany\\
  E-mail: \email{patrick.eckert@kip.uni-heidelberg.de}}

\abstract{We present a generic framework for the simulation of Silicon Photomultipliers (SiPMs) which enables detailed modelling of the SiPM response using basic SiPM parameters and geometry as an input. Depending on the specified SiPM properties which can be determined from basic characterisation measurements, the simulation generates the signal charge and pulse shape for arbitrary incident light pulse distributions.
The simulation has been validated in the whole dynamic range for a Hamamatsu S10362-11-100C MPPC and was used to study the effect of different noise sources like optical cross-talk and after-pulsing on the response curve and the photon-counting resolution.}


\begin{document}

\section{Introduction}
Silicon Photomultipliers (SiPMs) are a modern type of solid state photo-detectors which are suitable for a large range of applications such as calorimetry in high-energy physics, astrophysics or medical imaging \cite{Andreev:2006mb,Biland:2008zzb,moehrs2006detector}. The key features of SiPMs are a very compact size (typ.\,$1\,\mathrm{mm^{2}}$), insensitivity to magnetic fields, large gain, high photon detection efficiency and excellent single photon resolution.
\newline
A detailed understanding of the response to photons and the dependence on the operation parameters is essential for the utilisation of a photo-sensor. For SiPMs, the response depends on a multitude of effects like the photon detection efficiency, dark-rate, cross-talk, after-pulsing and pixel recovery time, as well as the spatial and time distribution of the incident light pulse. Due to the complex composition of the signal, an analytical description of the SiPM response is practically impossible. For this reason the Monte Carlo simulation framework GosSiP\footnote{\textbf{G}eneric framew\textbf{o}rk for the \textbf{s}imulation of \textbf{Si}licon \textbf{P}hotomultipliers} was developed. It provides a detailed SiPM model which allows to predict the response under arbitrary operating conditions and offers the opportunity to study the interplay of the different effects influencing the SiPM signal; in particular, it enables to identify the signal component which limits the performance for a specific application.
\newline
Earlier simulation studies presented in~\cite{Vacheret:2011zza,SanchezMajos:2008te} already show a good description of the SiPM response for low light intensities.
The simulation framework presented in this paper complements these studies by offering a more general simulation tool which is validated in the full dynamic range including the saturation regime at high light intensities. The new framework is used to study the effect of different noise contributions on the SiPM response and the photon-counting resolution.

\section{SiPM Response Characteristics}

Silicon Photomultipliers consist of an array of typically several hundred up to several thousand micro-cells (pixels) per $1\,\mathrm{mm^{2}}$ which are connected to a common output. Each pixel consists of an avalanche photo-diode (APD) operated at a bias voltage $V_{bias}$ larger than the diode breakdown voltage $V_{break}$ \cite{golovin2004novel}. 
In this so-called \textit{Geiger mode} operation, the electrical field at the p-n junction is large enough for free charge carriers to initiate a self-sustaining avalanche process (\textit{avalanche breakdown}). The avalanche is stopped with a large built-in quench resistor (passive quenching) or with an active electronic circuit controlling the bias voltage (active quenching). The presented SiPM simulation is designed for sensors with passive quenching, since this quenching scheme is implemented in most commercially available SiPMs. The following discussions are therefore restrained to SiPMs with passive quenching.
\newline
A consequence of the Geiger mode operation is that the charge produced in the avalanche process is independent of the number of incident photons hitting a pixel. The photon-counting capability of SiPMs is hence achieved by segmenting the detector area into an array of individual pixels. The number of detected photons can then be determined from the number of pixels where an avalanche breakdown occurred. Besides the photon detection, there are however several other effects which contribute to the signal and thus have to be taken into account when analysing the SiPM response:
\begin{itemize}

\item \textbf{Gain:}
The gain of the SiPM determines the charge which is produced by a single avalanche. In good approximation, the gain depends linearly on the pixel capacitance $C_{pixel}$ and the applied \textit{over-voltage} $V_{over}$ which is defined as the difference of the bias voltage and the breakdown voltage: 
\begin{equation}
	G = \frac{C_{pixel}}{q_e} \cdot V_{over} = \frac{C_{pixel}}{q_e} \cdot (V_{bias} - V_{break}) \mathrm{.}
	\label{eqn:gain}
\end{equation}
Here, $q_e$ is the elementary charge and $C_{pixel}$ the pixel capacitance.

\item \textbf{Photon Detection Efficiency:}
An essential property of every photo-detector is the photon detection efficiency (PDE) which describes the probability to detect a single photon. For SiPMs, this is determined by the quantum efficiency, the geometrical fill factor, surface reflections and the probability to initiate a Geiger discharge. The PDE significantly contributes to the photon-counting resolution due to the fluctuations in the number of detected photons.

\item \textbf{Dynamic Range \& Recovery Time:}
The finite number of pixels intrinsically limits the number of photons which can be detected simultaneously. This leads to a saturation behaviour of the SiPM response and fundamentally limits the dynamic range of the SiPM as well as the photon-counting resolution for high light intensities. For an ideal sensor and an infinitely short light pulse, the SiPM response can be described as
\begin{equation}
	N_{av}=N_{pixels} \cdot (1-e^{-\frac{\varepsilon_{PDE} \cdot N_{\gamma}}{N_{pixels}}}) \mathrm{,}
	\label{eqn:responseIdeal}
\end{equation}
where $N_{av}$ is the number of avalanche breakdowns, $N_{pixels}$ is the total number of pixels, $N_{\gamma}$ is the number if incident photons and $\varepsilon_{PDE}$ is the photon detection efficiency.
Considering photons which arrive within a certain time interval, the pixel recovery time can significantly influence the SiPM response. During the avalanche and quenching process the bias voltage at a pixel drops down to the breakdown voltage and recovers to the nominal operating voltage with a characteristic recovery time $\tau_{recovery}$:
\begin{equation}
  		V_{over}(t)=V_{over}(0) \cdot (1-e^{-t/\tau_{recovery}}) \mathrm{.}
  		\label{eqn:recoveryV}
\end{equation}
This recovery effect reduces the amount of charge which is produced in a subsequent avalanche breakdown.
In first approximation\footnote{Valid for small input impedance of the readout electronics} , the recovery time is determined by the pixel capacitance and the quench resistor:
\begin{equation}
	\tau_{recovery}=R_{quench} \cdot C_{pixel} \mathrm{.}
	\label{eqn:recoveryTime}
\end{equation}

\item \textbf{Cross-talk:}
It is well known that charge carriers crossing a p-n junction with reverse bias voltage beyond breakdown can create photons in the visible range \cite{newman1955visible}. The probability to emit a photon with sufficient energy to create an electron-hole pair is about $3 \times 10^{-5}$ per charge carrier crossing the junction. For a SiPM with a typical gain of $10^{6}$ about $30$ photons are produced per avalanche. These photons can propagate into nearby pixels and initiate a secondary avalanche breakdown faking a "real" photon hit. This effect is referred to as \textit{optical cross-talk} and degrades the photon-counting resolution of a SiPM due to the fluctuations in the number of cross-talk events. In addition, the increased pixel occupancy caused by optical cross-talk reduces the dynamic range of the sensor.

\item \textbf{After-pulses:}
During an avalanche, charge carriers can be trapped in the silicon lattice and be released after a characteristic time. This can cause a delayed second avalanche breakdown faking a photon signal and thus degrading the photon-counting resolution. In principle there can be several kinds of trapping centres with different characteristic trapping times. The data shown in section~\ref{sec:timeSpec} and \cite{Du:2008zzc} indicate the existence of two kinds of trapping centres, one with a slow and one with a fast time constant. Since cross-talk and after-pulse events are both triggered by an avalanche breakdown, these effects are often labelled as \textit{correlated noise}.

\item \textbf{Dark-rate:}
Avalanche breakdowns cannot only be triggered by photon absorption but also by thermal or field mediated excitations of electrons in the silicon lattice. These events will in the following be denoted as \textit{thermal pulses}. The thermal pulses and secondary avalanches caused by cross-talk and after-pulses make up the dark-rate of a sensor.

\item \textbf{Electronic Noise \& Excess Noise:}
The single photon resolution is determined by the pedestal noise $\sigma_{ped}$ and gain fluctuations $\sigma_{gain}$:
\begin{equation}
	\sigma(i)^{2}=\sigma_{ped}^{2}+i\cdot\sigma_{gain}^{2} \mathrm{,}
	\label{eqn:singlePhotonResolution}
\end{equation}
where $i$ is the number of avalanche breakdowns. The pedestal noise is defined by the signal fluctuation when no avalanche breakdown occurs. These fluctuations arise from leakage current and noise from the readout electronics. The gain fluctuations originate from pixel-to-pixel variations of the gain due to variations in the quenching resistance and fluctuations in the avalanche process. The value $\sigma_{gain}/G$ is often referred to as \textit{Excess Noise Factor}. In the following, $\sigma_{ped}$ and $\sigma_{gain}$ will be denoted as \textit{Electronic Noise} and \textit{Excess Noise}, respectively.
\end{itemize}

\section{SiPM Simulation Framework}
The prediction of the SiPM response for arbitrary operation conditions is non-trivial, due to the complex composition of the SiPM response -- especially in the case of large correlated noise, high light intensities and arbitrary light pulse shapes. For this reason, the Monte Carlo simulation framework \textit{GosSiP}\footnote{www.kip.uni-heidelberg.de/hep-detektoren/gossip}, which provides a detailed model of the SiPM response, has been developed. The simulation can be adjusted to any type of sensor by providing the basic SiPM parameters and pixel arrangements. In the following sections, this simulation framework and its comparison to data is presented.

The simulation framework is divided into three main parts and a graphical user interface which allows to adjust the SiPM and light source parameters, select a measurement and display the simulation result. Due to the modular structure the integration of GosSiP into other software frameworks, like e.\ g.\ Geant4\footnote{http://geant4.cern.ch}, is straightforward.
\begin{itemize}
  \item \textbf{Light source:}
  A basic simulation of a light source is implemented which generates a customisable light pulse illuminating the SiPM. The adjustable parameters are (i) the number of photons with the option of a Poisson smearing, (ii) the duration and time structure of the pulse (flat, Gaussian or exponential decay) and (iii) the position and shape (rectangular or elliptic) of the light spot. The position and time stamp of the individual photons within the pulse are randomly distributed according to the specified time and spatial structure. 
  
  \item \textbf{SiPM response:}
  The centrepiece of the GosSiP framework is the simulation of the signal charge and waveform for the specified SiPM parameters and light pulse input. All SiPM parameters can be customised in order to model different types of SiPMs.

  \item \textbf{Data acquisition:}
  A simple simulation of the basic characterisation measurements using standard electronics is provided, including measurements of the charge and time spectra, dark-rate threshold scans and response curve determination.
\end{itemize}
In the first simulation step, a list of incoming photons is generated by the internal light source simulation. Alternatively, this list can also be generated by external simulation tools. The photon information is then passed to the SiPM simulation which models the response of the sensor to the incident light pulse. Each photon hitting the sensor can trigger an avalanche with a certain probability given by the PDE. To simulate the thermal noise, random pulses are generated according to the specified thermal pulse time constant (see equation~\ref{eqn:Ptp}). The information about the initial time of each avalanche as well as the position of the fired pixel is stored in a list which is then processed chronologically according to the following steps.

The charge generated by the avalanche is determined using the specified gain and a Gaussian smearing to account for the excess noise. Then cross-talk and after-pulse events are generated according to the specified cross-talk and after-pulse probabilities (see equation~\ref{eqn:Pap} \&~\ref{eqn:Pxt}) and inserted chronologically into the avalanche list. This naturally enables higher-order noise cascades from an initial avalanche.
A simple cross-talk model is used in which cross-talk can only occur in the four neighbouring pixels. This model yields a good description of the data as shown in earlier publications \cite{SanchezMajos:2008te,Vacheret:2011zza} and validated below. The after-pulse model is based on the probability density function given by equation~\ref{eqn:Pap} which depends on a characteristic trapping time constant and a trapping probability. Two independent after-pulse components are implemented accounting for a slow and a fast trapping time constant.
\newline
The chronological processing of the avalanche list is essential in order to model the pixel recovery effects. The recovery process influences all effects which contribute to the signal (i.\,e.\,gain, excess noise, PDE, thermal rate, cross-talk, after-pulses) due to the changing over-voltage (see equation~\ref{eqn:recoveryV}).
The recovery of the gain is modelled using equation~\ref{eqn:recoveryV} and~\ref{eqn:gain}:
\begin{equation}
        G(t)=G(0) \cdot (1-e^{-t/\tau_{recovery}}) \mathrm{.}
		\label{eqn:recoveryG}
\end{equation}
The recovery of the PDE, excess noise, cross-talk, after-pulse and thermal pulse probabilities is modelled in the same way. This is only a rough approximation since the latter effects in general do not depend linearly on the over-voltage, as it is the case for the gain. However, the impact of this approximation is small due to the reduced signal amplitude of a firing pixel which is not fully recovered. A more sophisticated recovery model can be implemented easily. This would however necessitate a more complex simulation input, since the over-voltage dependence of all effects would have to be provided.
\newline
In the final step of the simulation chain, the simulated output signal can be processed by the simulation of the data acquisition which models the most common characterisation measurements. Alternatively, the simulated output can be passed to an external software for a custom digitisation of the signal.

\section{Simulation Input \& Validation}
\label{sec:inputandvalidation}

\subsection{Input Parameter Determination}
\label{sec:characterisation}
The simulation input is given by the following basic SiPM parameters: PDE, gain, excess noise, electronic noise, thermal noise time constant, cross-talk probability, after-pulse probabilities and after-pulse time constants, recovery time, number of pixels, the pixel arrangement and the single pixel pulse shape. These parameters can be determined from basic characterisation measurements which will be briefly described in this section. A detailed description of the measurement setup and method can be found in \cite{Eckert:2010gs}.
The following study was carried out using a Hamamatsu MPPC\footnote{Multi Pixel Photon Counter} S10362-11-100C device with 100 pixels, operated at an over-voltage of 0.5\,V and 1.0\,V. This covers the range of low and high noise operation of the sensor. Table~\ref{tab:inputParas} summarises the obtained values for the input parameters.

\newcolumntype{C}{>{\centering\arraybackslash}X}
\begin{table}[tb]
\centering
	\caption{Simulation input parameters determined by the characterisation measurements. The PDE is measured at $\lambda \approx 658\,nm$ (red laser diode), whereas the maximum sensitivity of the used sensor is at $\lambda \approx 460\,nm$ (see \cite{Eckert:2010gs}).}
	\begin{tabularx}{.75\linewidth}{lCC}
		\toprule
		Parameter											& $V_{over} = 0.5\,V$ 	& $V_{over} = 1.0\,V$		\\
		\hline \hline
		$PDE\,[\%]$										& $4.8 \pm 0.3$				& $11.6 \pm 0.6$			\\
		$Gain\,[q_e\cdot10^{5}]$						& $9.2 \pm 0.2$				& $21.5 \pm 0.2$			\\
		$\sigma_{gain}/\mathrm{Gain}\,[\%]$	& $13.2 \pm 0.3$			& $5.8 \pm 0.3$				\\
		$\sigma_{ped}/\mathrm{Gain}\,[\%]$	& $6.4 \pm 0.3$				& $9.1 \pm 0.3$				\\
		$\tau_{DR}\,[ns]$								& $6632 \pm 66$			& $2821 \pm 28$			\\
		$\tau_{AP_{slow}}\,[ns]$						& $211 \pm 26$				& $181 \pm 6$				\\
		$\tau_{AP_{fast}}\,[ns]$						& $61 \pm 7$					& $61 \pm 2$					\\
		$P_{AP_{slow}}\,[\%]$						& $2.3 \pm 0.1$				& $13.0 \pm 0.6$			\\
		$P_{AP_{fast}}\,[\%]$							& $3.6 \pm 0.2$				& $16.3 \pm 0.4$			\\
		$P_{CT,n\geq1}\,[\%]$								& $2.4 \pm 0.1$				& $14.3 \pm 0.6$			\\
		$\tau_{recovery}\,[ns]$						& \multicolumn{2}{c}{$38.4 \pm 1.2$}					\\
		\bottomrule
	\end{tabularx}
	\label{tab:inputParas}
\end{table}

\begin{figure}[htb]
\centering
	\includegraphics[width=.75\linewidth]{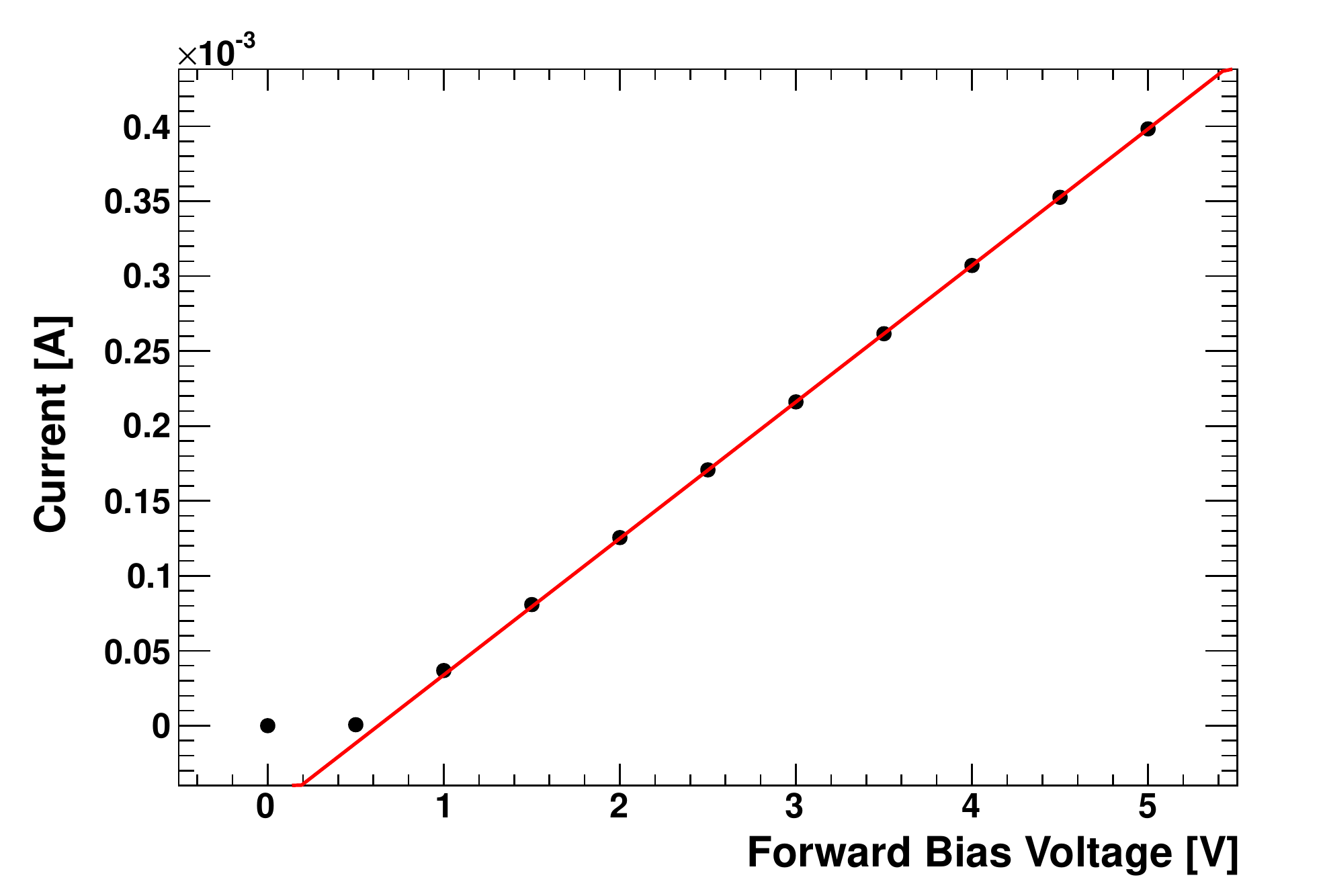}
	\caption{IV--Curve: The quench resistance determined from a linear fit to the the data above 1\,V forward bias voltage as in this region the resistance of the pn-junction is negligible.}
	\label{fig:IVcurve}
\end{figure}

\paragraph{Quench Resistance \& Recovery Time}
The quench resistance $R_{quench}$ is determined from a linear fit of the I--V curve for forward bias operation (see figure~\ref{fig:IVcurve}):
\begin{equation}
	R = R_{quench}/N_{pixels} + R_{OC} = \Delta V / \Delta I \mathrm{,}
\end{equation}
where $R_{OC}$ is a serial resistance in the operation circuit.
For the tested sensor a value of $R_{quench} = 96 \pm 3 \mathrm{\,k\Omega}$ is obtained.
This yields an indirect measurement of the pixel recovery time using equation~\ref{eqn:recoveryTime} and a pixel capacitance of $C_{pixel} = 400 \pm 1\,\mathrm{fF}$ which is determined from the gain -- voltage dependence (see section~\ref{sec:chargeSpectrum}) The resulting recovery time is $\tau_{recovery} = 38.4 \pm 1.2 \mathrm{\,ns}$ which is comparable to the value of $\thicksim33\mathrm{\,ns}$ presented in a study \cite{Oide:2006zz} based on a waveform analysis.

\paragraph{Charge Spectrum}
\label{sec:chargeSpectrum}
\begin{figure}[htb]
\centering
	\includegraphics[width=.75\linewidth]{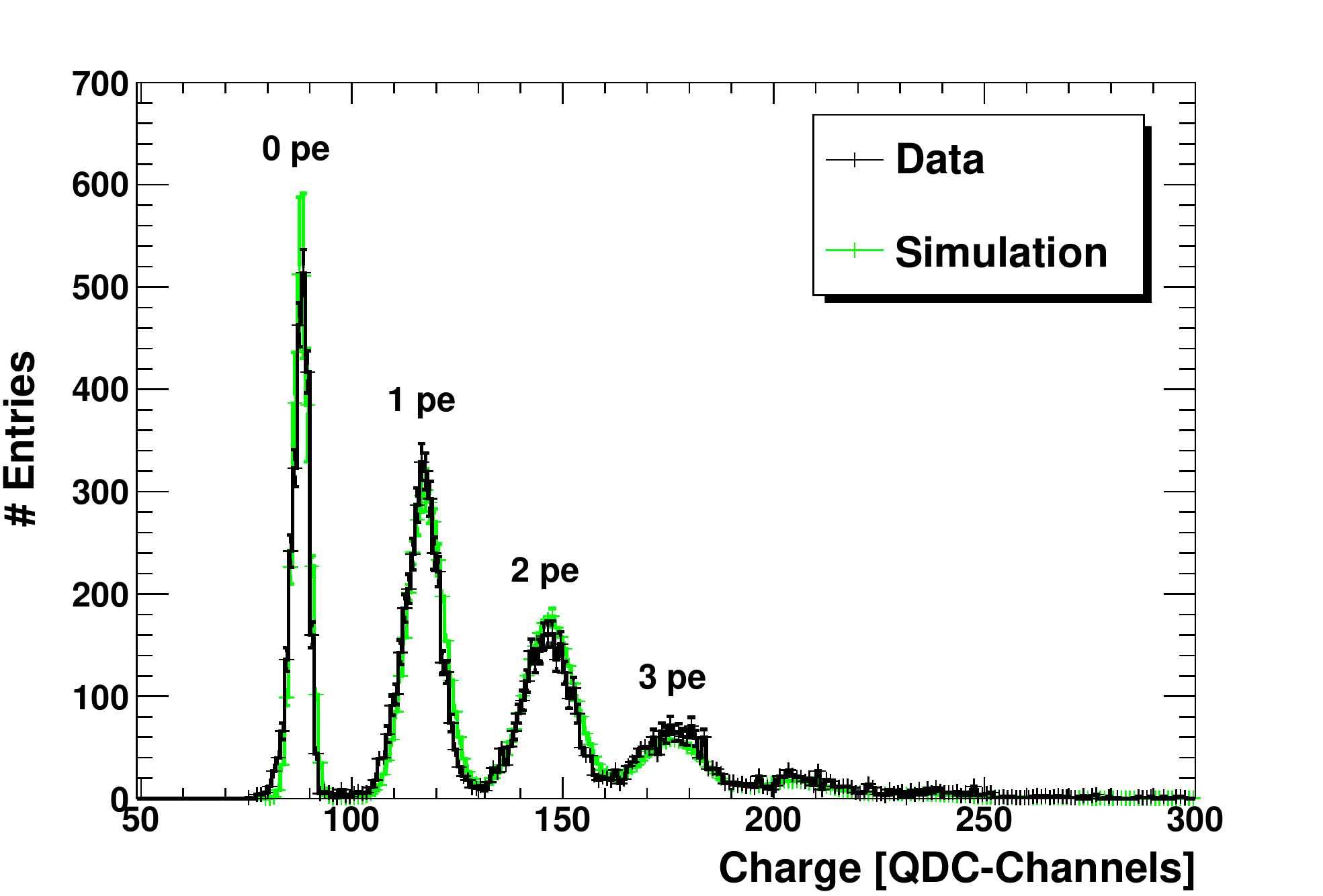}
	\caption{Single photo-electron spectrum for 0.5\,V over-voltage. Each peak corresponds to a certain number of photo-electrons (pe).}
	\label{fig:QDCSpec}
\end{figure}
The charge spectrum is one of the most important tools for the characterisation of SiPMs, since it provides information about a multitude of parameters. It is acquired by illuminating the sensor with a short light pulse ($\mathcal{O}(\text{ns})$) and integrating the signal charge within a certain time window using a QDC\footnote{Charge-to-Digital converter; LeCroy Model 2249, $50\,\mathrm{\Omega}$ input impedance.}. A typical charge spectrum for low light levels (\textit{single pixel spectrum}), as displayed in figure~\ref{fig:QDCSpec}, shows several well separated peaks which correspond to a certain number of avalanche breakdowns. All events with no pixel signals are contained in the first peak (pedestal), the second peak contains all events with one pixel firing, and so on. The SiPM gain is given by the charge which is produced by a single avalanche and hence is equivalent to the charge difference between two peaks. The voltage dependence of the gain described by equation~\ref{eqn:gain} can be used to determine the pixel capacitance and breakdown voltage.\newline
The number of photons in the incident light pulse, and therefore also the number of detected photons, is expected to be Poisson distributed. However, the distribution of the number of detected photons is distorted due to cross-talk and after-pulses which fake photo-electron signals.
The mean value of the charge spectrum therefore corresponds to the mean number of avalanche breakdowns $N_{av}$ including signals from cross-talk, after-pulses and thermal pulses
\begin{equation}
	N_{av} = (Q-Q_{ped})/G \mathrm{,}
	\label{eqn:nFired}
\end{equation}
with $Q$ denoting the mean value of the charge spectrum and $Q_{ped}$ denoting the pedestal value. Exploiting Poisson statistics, the mean number of detected photons $N_{pe}$ without any bias from noise can be determined from the number of pedestal events which are not affected by cross-talk and after-pulses:
\begin{align}
  		P(0,N_{pe}) &= e^{-N_{pe}} \\
  		\rightarrow N_{pe} &= - ln(P(0,N_{pe})) = -ln(\frac{c_{DR} \cdot N_{ped}}{N_{tot}}) \mathrm{.}
  		\label{eqn:nDetected}
\end{align}
Here, $N_{tot}$ is the total number of recorded events and $N_{ped}$ is the number of pedestal events. $c_{DR}$ is a correction factor rescaling the value of $N_{ped}$ in order to account for the dark-rate which reduce the number of pedestal events. This correction factor can be determined from a charge spectrum of the dark-rate \cite{Eckert:2010gs}:
\begin{equation}
	c_{DR} = N_{tot}^{DR}/N_{ped}^{DR} \mathrm{,}
\end{equation}
where $c_{DR}-1$ corresponds to the probability for a dark-rate event to occur during the charge integration.
\newline
If the charge spectrum is measured using a referenced light source, this statistical analysis allows to determine the PDE without the effects of cross-talk, after-pulses and dark-rate. A detailed description of this measurement is given in \cite{Eckert:2010gs}.\newline
If no pixel fires only the charge generated by electronic noise is integrated; this is reflected in the width of the pedestal peak $\sigma_{ped}$.
The excess noise is given by the additional broadening of the 1\,pe peak compared to the pedestal:
\begin{equation}
	\sigma_{gain} = \sqrt{\sigma_{1pe}^{2}-\sigma_{ped}^{2}} \mathrm{,}
\end{equation}
where $\sigma_{1pe}$ is the width of the 1\,pe peak.

\paragraph{Dark-rate Time Spectrum}
\label{sec:timeSpec}
The thermal pulse and after-pulse characteristics can be determined from a statistical analysis of the time interval between two consecutive dark-rate pulses.
The spectrum of this time interval is acquired using a discriminator at $0.5\,\mathrm{pe}$ level and measuring the time stamps of the discriminator pulses with a TDC\footnote{Time-to-Digital Converter; CAEN V1290A} module. For our setup, this method allows to stably acquire time intervals down to $t \approx 100\,\mathrm{ns}$, due to the relatively wide pulse shape of the signal.
The time spectrum represents the probability density for a dark-rate pulse to occur after a time $t$ with respect to the previous pulse.
The probability density for the thermal pulse component is given by:
\begin{equation}
	p_{TP}(t) = \frac{1}{\tau_{TP}} \cdot e^{-\frac{t}{\tau_{TP}}} \mathrm{,}
	\label{eqn:Ptp}
\end{equation}
where $\tau_{TP}$ is the characteristic thermal pulse time constant. For after-pulses with a time constant of $\tau_{AP}$, this probability is multiplied with the after-pulse probability $P_{AP}$:
\begin{equation}
	p_{AP}(t) = \frac{P_{AP}}{\tau_{AP}} \cdot e^{-\frac{t}{\tau_{AP}}} \mathrm{.}
	\label{eqn:Pap}
\end{equation}
The measured time spectrum can be approximated with a combination of the thermal pulse, slow and fast after-pulse component:
\begin{align}
\label{eqn:timeSpec}
	&N(t) = N \cdot \\
	&\left( p_{TP}(t) \cdot [1-\int_0^t p_{AP_s}(t^\prime)\,\mathrm{dt^\prime}] \cdot [1-\int_0^t p_{AP_f}(t^\prime)\,\mathrm{dt^\prime}] \right. \nonumber \\
	&+ p_{AP_s}(t) \cdot [1-\int_0^t p_{TP}(t^\prime)\,\mathrm{dt^\prime}] \cdot [1-\int_0^t p_{AP_f}(t^\prime)\,\mathrm{dt^\prime}] \nonumber \\
	&+ \left. p_{AP_f}(t) \cdot [1-\int_0^t p_{TP}(t^\prime)\,\mathrm{dt^\prime}] \cdot [1-\int_0^t p_{AP_f}(t^\prime)\,\mathrm{dt^\prime}] \right) \nonumber
\end{align}
where $p_{AP_{s/f}}(t)$ denotes the probability density for the slow and fast after-pulse component, respectively.
The simulation input parameters $\tau_{TP},\tau_{AP_s},\tau_{AP_f},P_{AP_s} \mathrm{and\,} P_{AP_f}$ can be extracted by fitting the measured time spectrum with formula~\ref{eqn:timeSpec}.
However, formula~\ref{eqn:timeSpec} is a first order approximation and does not take into account cross-talk and pixel recovery which influence the number of after-pulse events. Therefore, the values for the time constants and after-pulse probability determined by the fit exhibit a systematic uncertainty. The minimum time interval of $t \approx 100\,\mathrm{ns}$ which can be achieved with our setup also limits the sensitivity to fast after-pulses which may lead to systematic errors in the fit.
In order to account for these effects, the SiPM simulation output is fitted to the data by iterative variation of the input parameters. The resulting values for the input parameters show a deviation of up to $20\%$ compared to the values obtained from the analytical fit using equation~\ref{eqn:timeSpec}.
With a more sophisticated setup for the determination of the after-pulse properties (e.\,g.\,\cite{Vacheret:2011zza,Du:2008zzc}) it should be possible to determine the simulation input parameters via an analytical fit similar to equation~\ref{eqn:timeSpec}.
\begin{figure}[tb]
\centering
	\includegraphics[width=.75\linewidth]{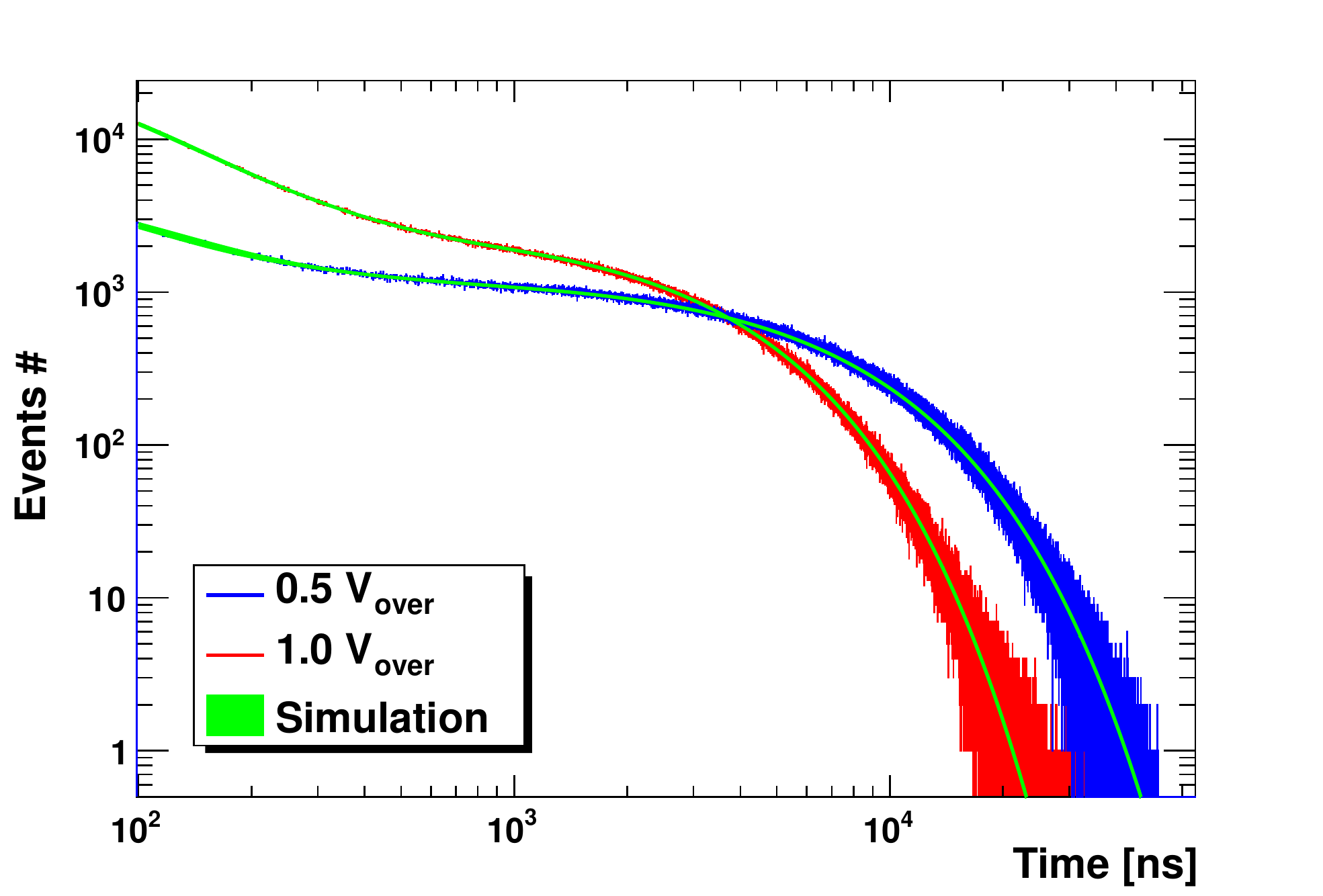}
	\caption{Spectrum of the time intervals between two consecutive dark-rate pulses.}
	\label{fig:timeSpec}
\end{figure}

\paragraph{Dark-rate Threshold Scan}
\label{sec:threshScan}
\begin{figure}[htb]
\centering
	\includegraphics[width=.75\linewidth]{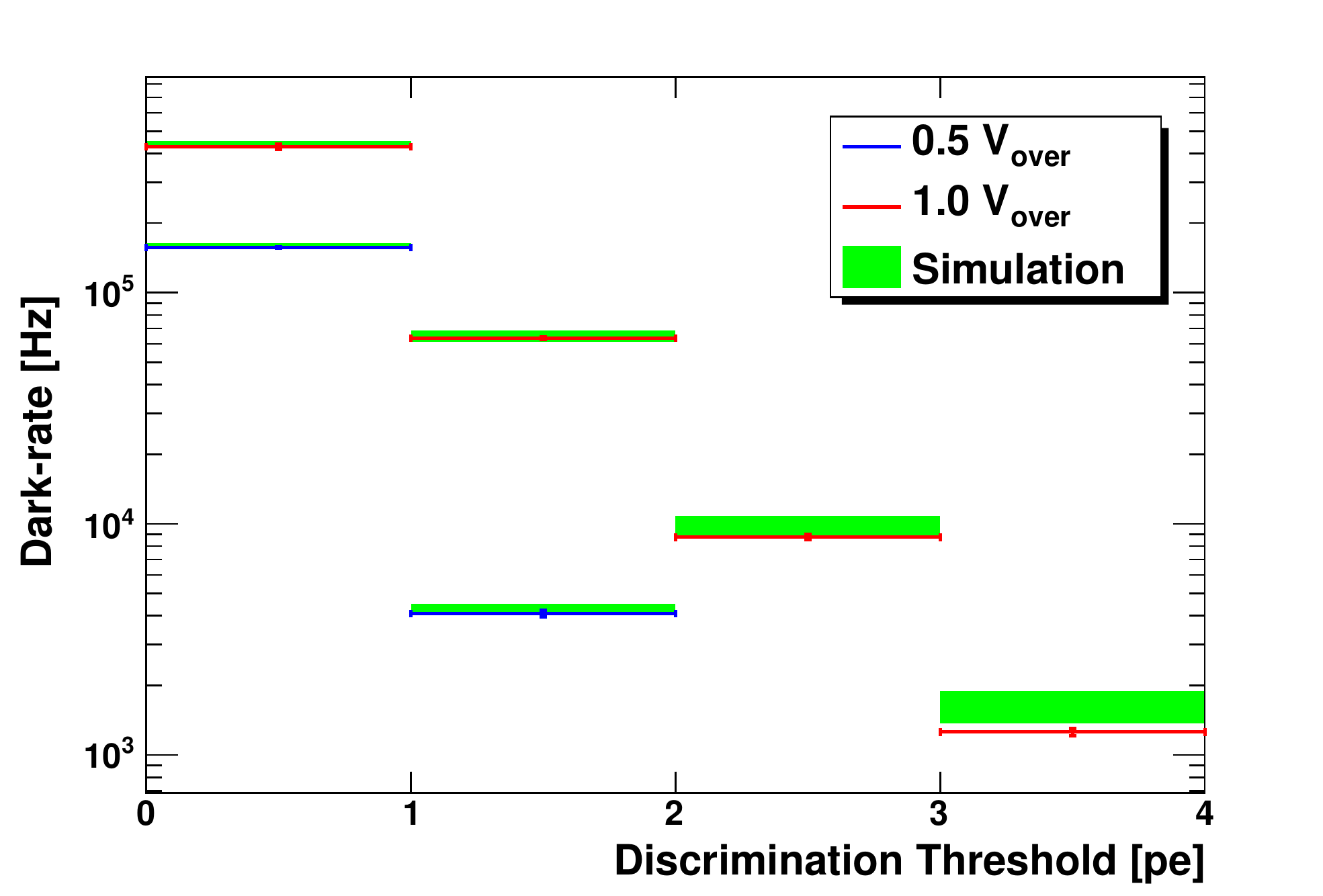}
	\caption{Dark-rate measured at different discrimination thresholds.}
  	\label{fig:threshScan}
\end{figure}
The cross-talk probability can be determined by counting the number of dark-rate events with a certain number of fired pixels. This is done using a discriminator module which selects dark-rate pulses with a certain amplitude. The discriminator signals are then counted with a scaler module\footnote{LeCroy Model 2250B}. Figure~\ref{fig:threshScan} shows the dark-rate as a function of the applied discrimination threshold.
The total dark-rate at 0.5\,pe threshold is composed of the thermal pulse rate and a contribution from after-pulses.
The events where two or more pixels fire at the same time, which correspond to a discrimination threshold of 1.5\,pe, are caused by dark-rate events triggering optical cross-talk. The probability for two thermal pulses to occur at the same time is negligible.
The probability for an avalanche to generate one or more cross-talk events $P_{CT,n\geq1}$ is given by the ratio of the dark-rates at 1.5\,pe and 0.5\,pe threshold: $P_{CT,n\geq1}= DR_{1.5pe}/DR_{0.5pe}$; this is the cross-talk input parameter for the simulation.
A simplified cross-talk model is implemented in the simulation taking into account only the four directly neighbouring pixels which generate a cross-talk event with a respective probability of $P_{CT}$. This value is related to the input value via the following equation:
\begin{equation}
	(1-P_{CT})^4 = 1-P_{CT,n\geq1} \mathrm{.}
	\label{eqn:Pxt}
\end{equation}

\paragraph{Simulation Cross-checks}
\label{sec:crosscheck}
Comparing the simulated response to the characterisation measurements is important to validate the physics models used in the simulation. The shape of the simulated time spectrum (see figure~\ref{fig:timeSpec}) agrees well with the measurement confirming the implemented after-pulse and thermal pulse model. The simulation also reproduces the total dark-rate at 0.5\,pe threshold shown in figure~\ref{fig:threshScan} 
which independently validates the after-pulse and thermal pulse model. The rates at 2.5\,pe and 3.5\,pe threshold are slightly over-estimated by the simulation; this could be attributed to the simplified cross-talk model or the simplified discriminator model\footnote{The discriminator is modelled using a simple parametrization.} used in the simulation. However, the deviation is small and the probability $P_{CT,n\geq2}$ for two or more cross-talk events which corresponds to the rate at 2.5\,pe threshold is smaller than $2 \%$. Therefore, the effect can be safely neglected.
\newline
The simulated charge spectrum (see figure~\ref{fig:QDCSpec}) also agrees well with the data confirming the implemented gain, excess noise, and photon detection model. Since the amplitudes of the individual peaks in the spectrum are also influenced by cross-talk and after-pulses, this again shows the validity of the implementation of these two effects.
\newline
These cross-checks verify, that the simulation offers a good description of the SiPM dark-rate and the response to low light intensities. 
Similar results with a comparable simulation approach have been published in \cite{Vacheret:2011zza}. However, for large light intensities, the high pixel occupancy and pixel recovery have a significant influence on the response and effects like cross-talk, after-pulses and photo detection.
Therefore, the validation of the simulation was extended to cover the whole dynamic range. This will be discussed in the following chapter.

\subsection{Response Curve \& Resolution}
\label{sec:validation}
For low light intensities, the individual pixel signals can be treated as independent which allows a rather simple description of the SiPM response. This is not the case for high light intensities due to the saturation of the signal. In order to validate the simulation over the whole dynamic range including the low intensity region as well as the non-linear saturation region, the response curve and signal resolution were measured and compared to the simulation results.
The response curve is a sensitive quantity, since it is influenced by all SiPM parameters which enter the simulation.
\begin{figure}[ptb]
\centering
	\includegraphics[width=.75\linewidth]{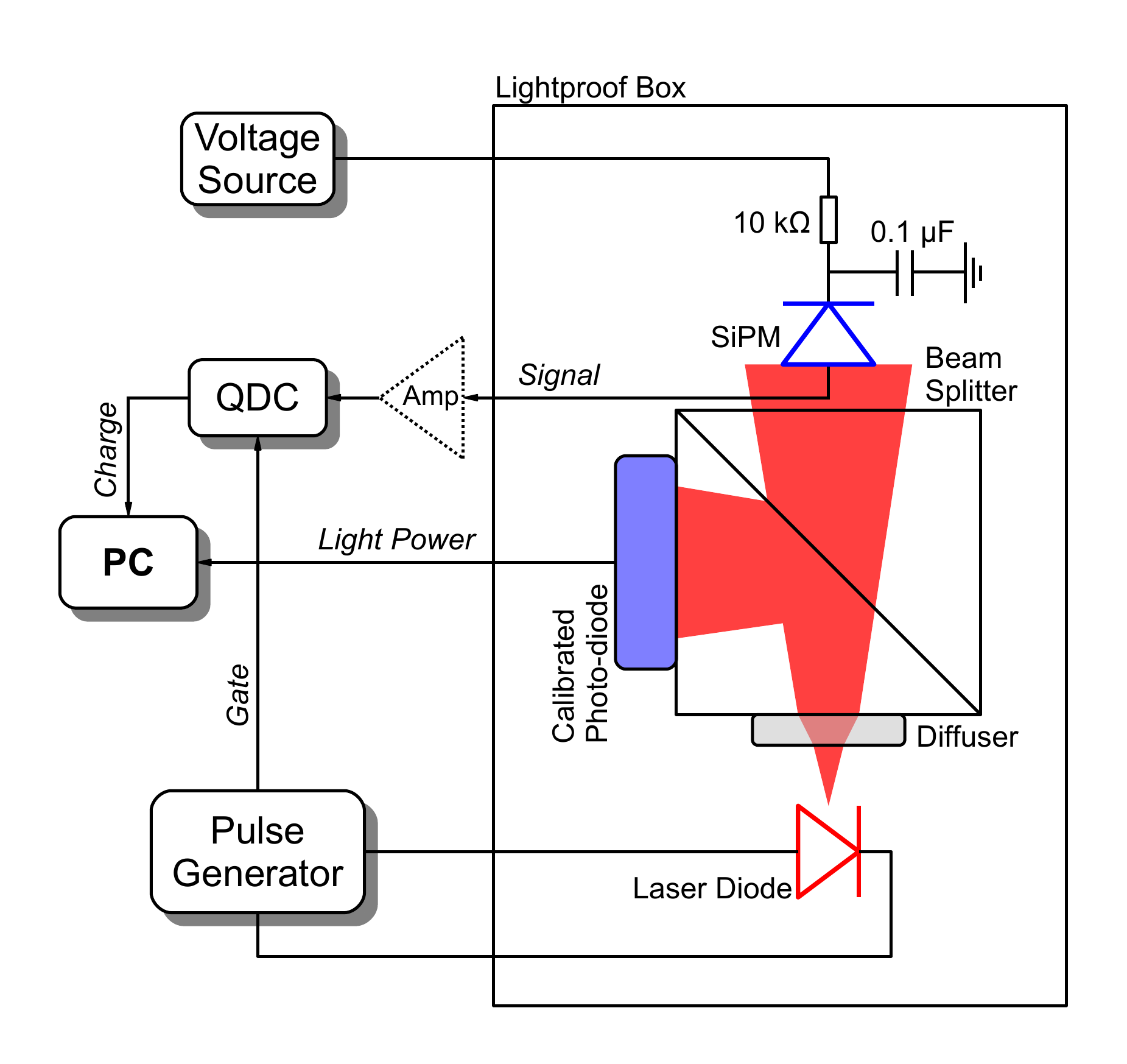}
	\caption{Schematic view of the setup for the measurement of the response curve.}
	\label{fig:setup}
\end{figure}

The basic principle of the measurement is to determine the SiPM response as a function of the incident photon flux (see figure~\ref{fig:setup}). The SiPM is illuminated using a laser diode emitting at a wavelength of $\lambda \approx 658\,\mathrm{nm}$, which is driven by a pulse generator with 4\,ns pulse width. A beam splitter evenly distributes the light to the SiPM and a NIST\footnote{National Institute of Standards and Technology} calibrated photo-diode which serves as a reference sensor with ideal linear response. The charge of the SiPM signal is measured with a QDC within an integration gate of 300\,ns.

\begin{figure}[htb]
\centering
	\includegraphics[width=.75\linewidth]{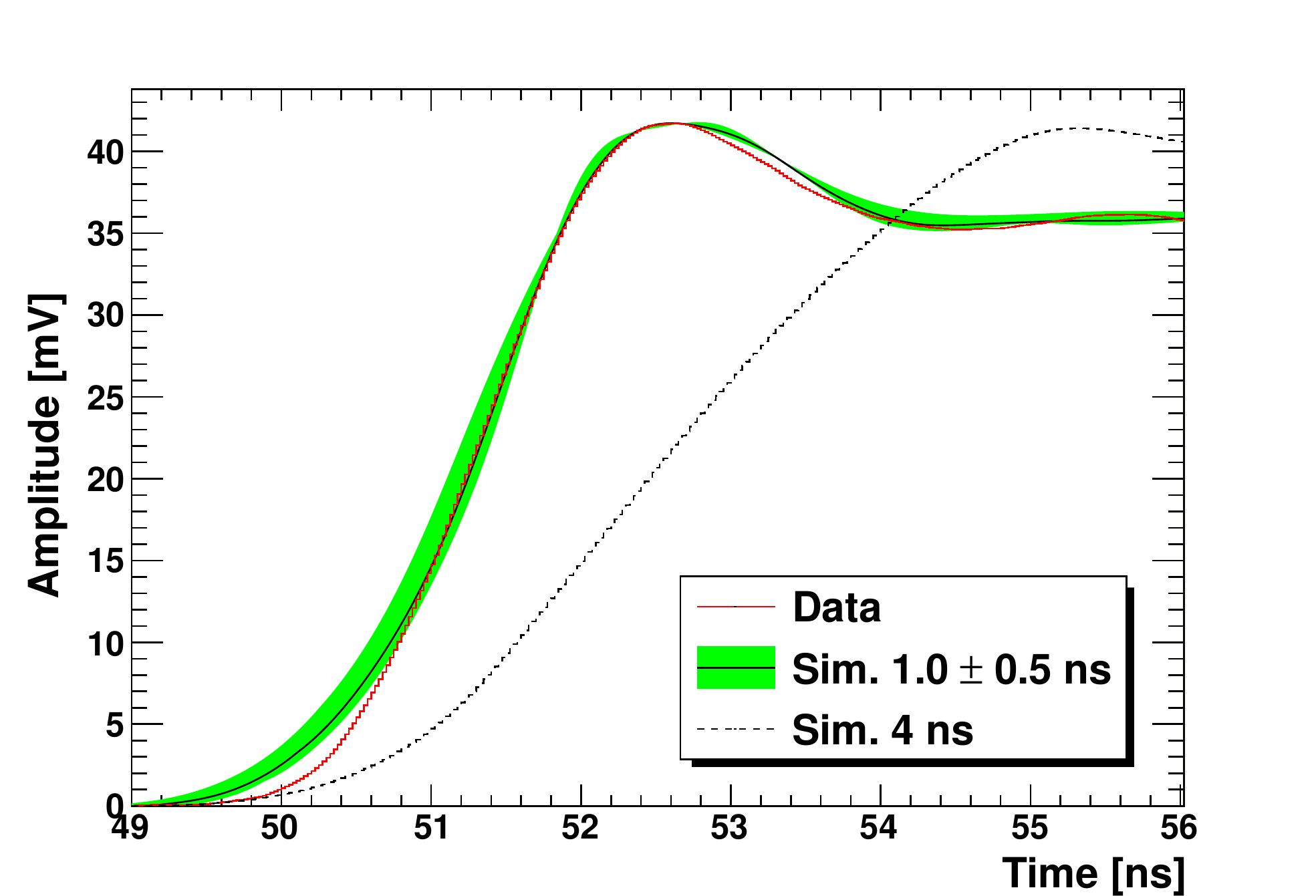}
	\caption{Leading edge of the signal (about 20 firing pixels) for data (red), reconstructed waveform assuming a flat $1.0 \pm 0.5\,\mathrm{ns}$ light pulse (black line and green uncertainty band), and reconstructed waveform assuming a flat $4.0\,\mathrm{ns}$ light pulse (black dashed line).}
	\label{fig:lightSpot}
\end{figure}

The light spot properties are crucial for the measurement of the response curve. In order to achieve a homogeneous illumination of the SiPM, a diffuser is used to obtain a uniform light spot. In addition, the MPPC S10362-11-100C sensor was chosen for this measurement due to its long pixel recovery time of $38.4\mathrm{\,ns}$ which minimizes the influence of the time structure of the light pulse.
However, it was observed, that the SiPM response in the non-linear range still depends weakly on the width of the voltage pulse driving the laser diode.
Since the time response of the laser diode to the voltage pulse is not known, the duration of the light pulse was estimated by analysing the leading edge of the SiPM signal at medium light intensities ($\approx 20$ firing pixels).
The signal waveform was reconstructed using the simulation adding up averaged 1\,pe waveforms which have been measured with an oscilloscope.
The best agreement of measured and reconstructed signal was achieved with a light pulse time interval of $\Delta t \approx 1.0 \pm 0.5\,\mathrm{ns}$ and a flat time distribution (see figure~\ref{fig:lightSpot}).
Assuming a pulse width of 4\,ns, which corresponds to the pulse width of the pulse generator, a significant deviation between measured and reconstructed signal is observed.

The number of photons, $N_{\gamma}$, hitting the sensor is determined with the NIST reference sensor. The sensor is calibrated using SiPM charge spectra at low intensity, from which $N_{\gamma}$ is obtained via the relation $N_{\gamma} = N_{pe}/PDE$. $N_{pe}$, the average number of photons detected with the SiPM, is extracted via equation~\ref{eqn:nDetected} and the PDE is taken from Table~\ref{tab:inputParas}. The uncertainty of this calibration procedure leads to the systematic uncertainty on the obtained value of $N_{\gamma}$.
\newline
The SiPM response, i.\,e.\ the number of fired pixels $N_{av}$, is determined from the measured signal charge using equation~\ref{eqn:nFired}. Due to the limited dynamic range of the QDC the response at high light intensities cannot be measured using pre-amplification. The necessary gain factor (in units of QDC-channels), $G$, for non-amplified signals thus has to be determined. This is done by comparing the measured charge $Q-Q_{ped}$ at low light levels with and without amplification. The ratio of the measured charges yields the amplification factor which is used to correct the gain measured using amplified single photon spectra. The uncertainty on this amplification factor leads to the dominant systematic error on $N_{av}$.

\begin{figure}[p]
\centering
	\includegraphics[width=.75\linewidth]{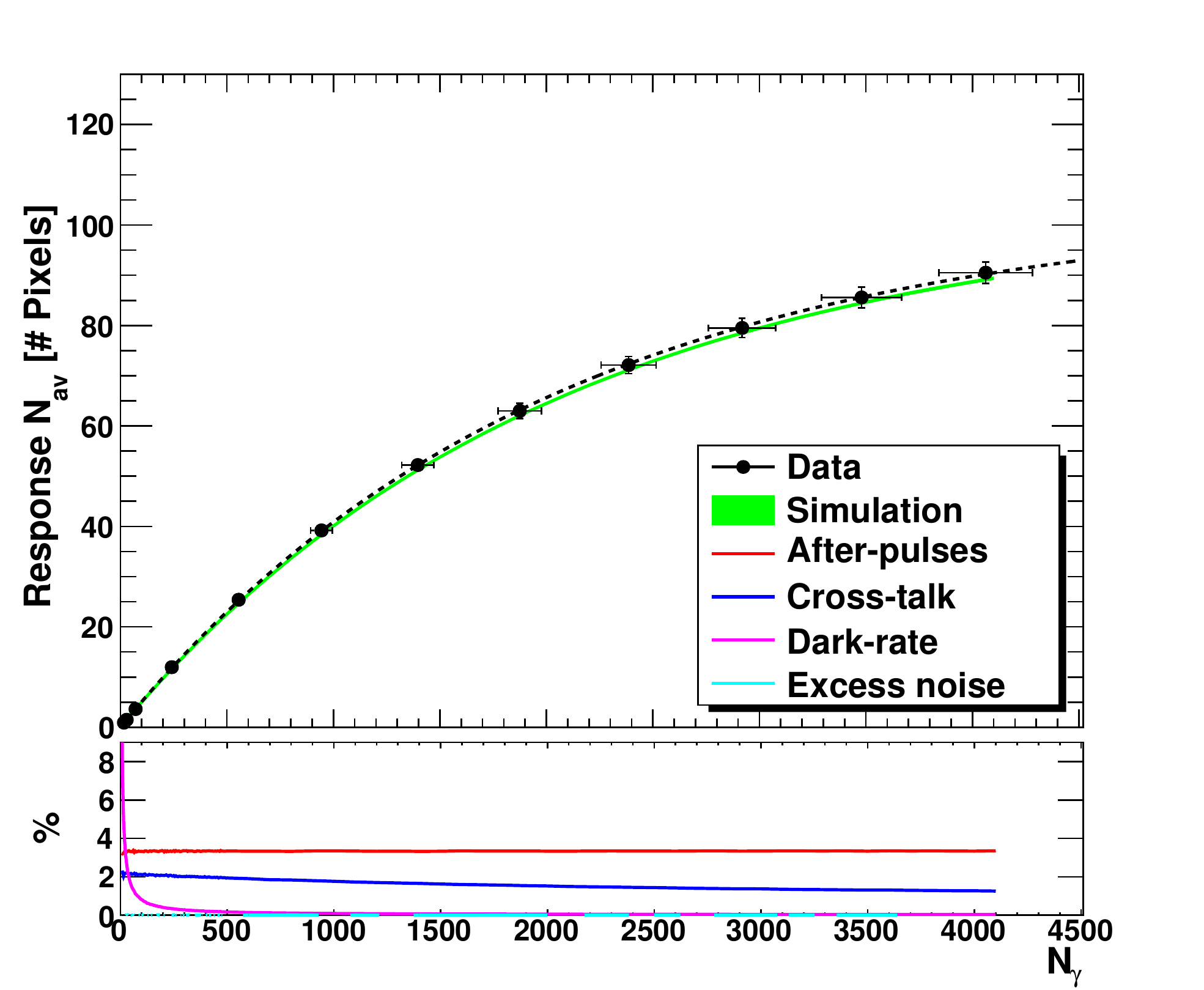}
	\caption{\textbf{Top:} SiPM response for low noise (0.5\,V over-voltage). \textbf{Bottom:} Contribution from the individual noise sources.}
	\label{fig:dynRange0.5}
	
	\includegraphics[width=.75\linewidth]{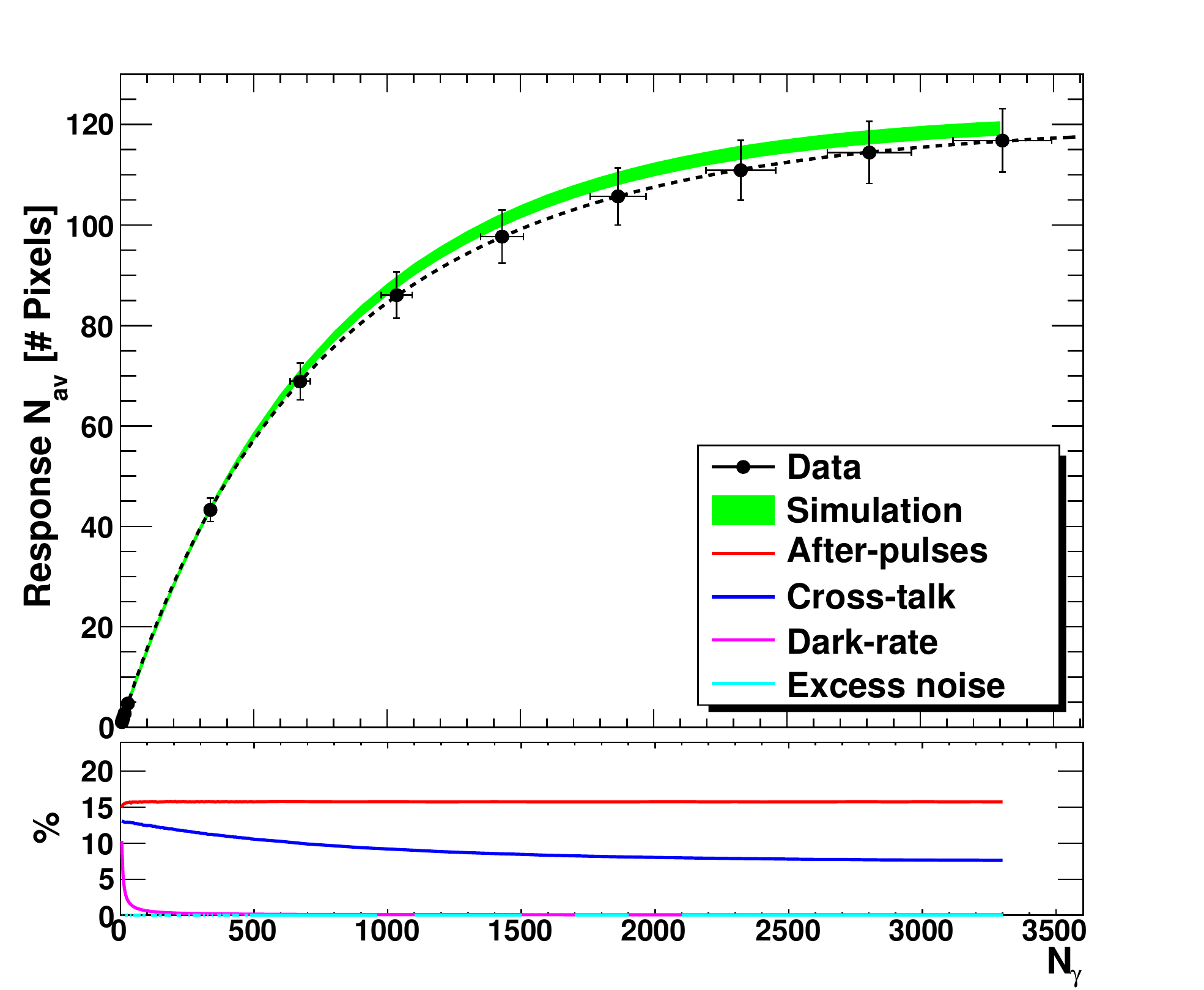}
	\caption{\textbf{Top:} SiPM response for high noise (1.0\,V over-voltage). \textbf{Bottom:} Contribution from the individual noise sources.}
	\label{fig:dynRange1.0}
\end{figure}
	
\begin{figure}[p]
\centering
	\includegraphics[width=.75\linewidth]{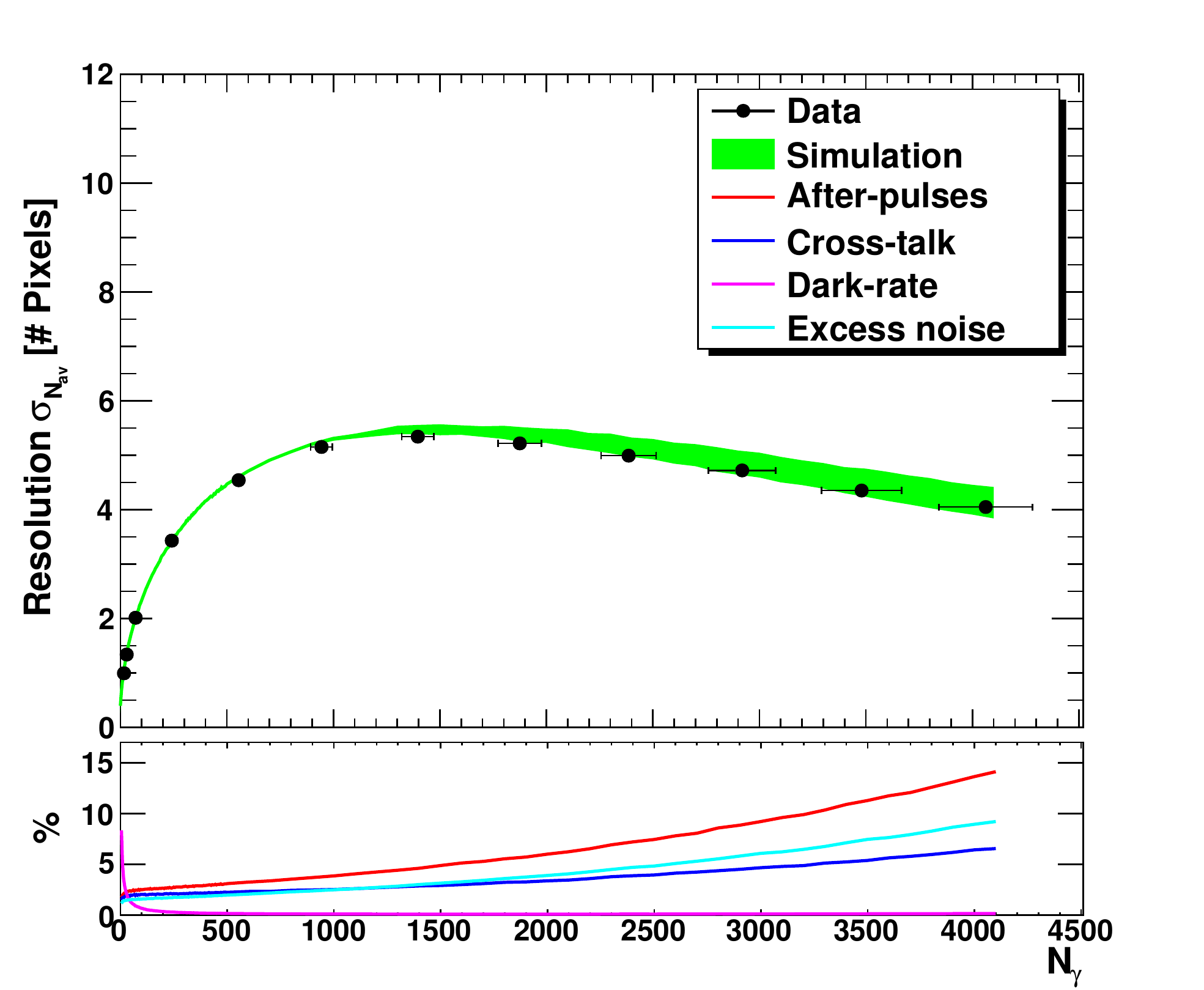}
	\caption{\textbf{Top:} Resolution of the response for low noise (0.5\,V over-voltage). \textbf{Bottom:} Contribution from the individual noise sources.}
	\label{fig:rms0.5}
	
	\includegraphics[width=.75\linewidth]{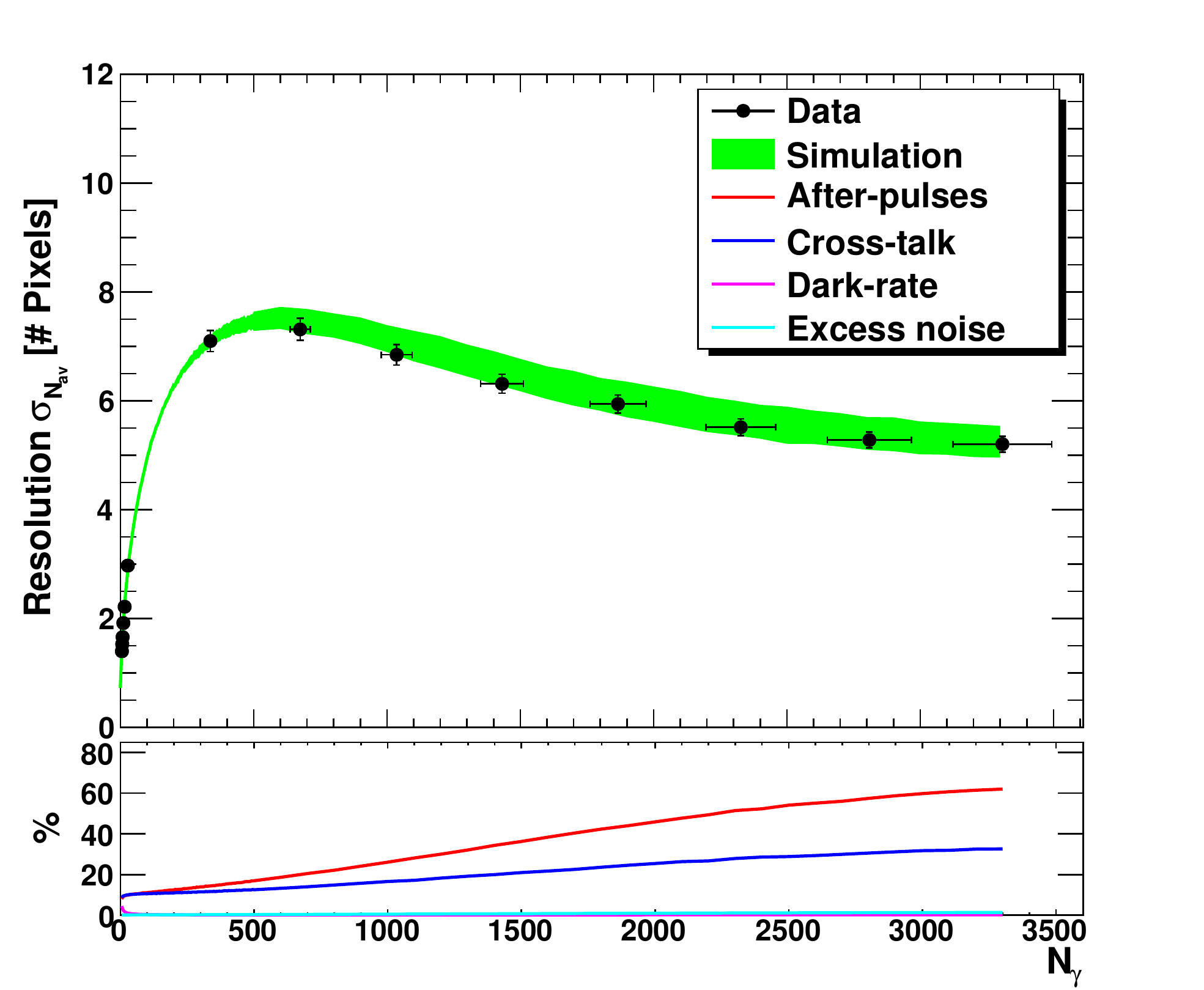}
	\caption{\textbf{Top:} Resolution of the response for high noise (1.0\,V over-voltage). \textbf{Bottom:} Contribution from the individual noise sources.}
	\label{fig:rms1.0}
\end{figure}

\begin{figure}[p]
\centering
	\includegraphics[width=.75\linewidth]{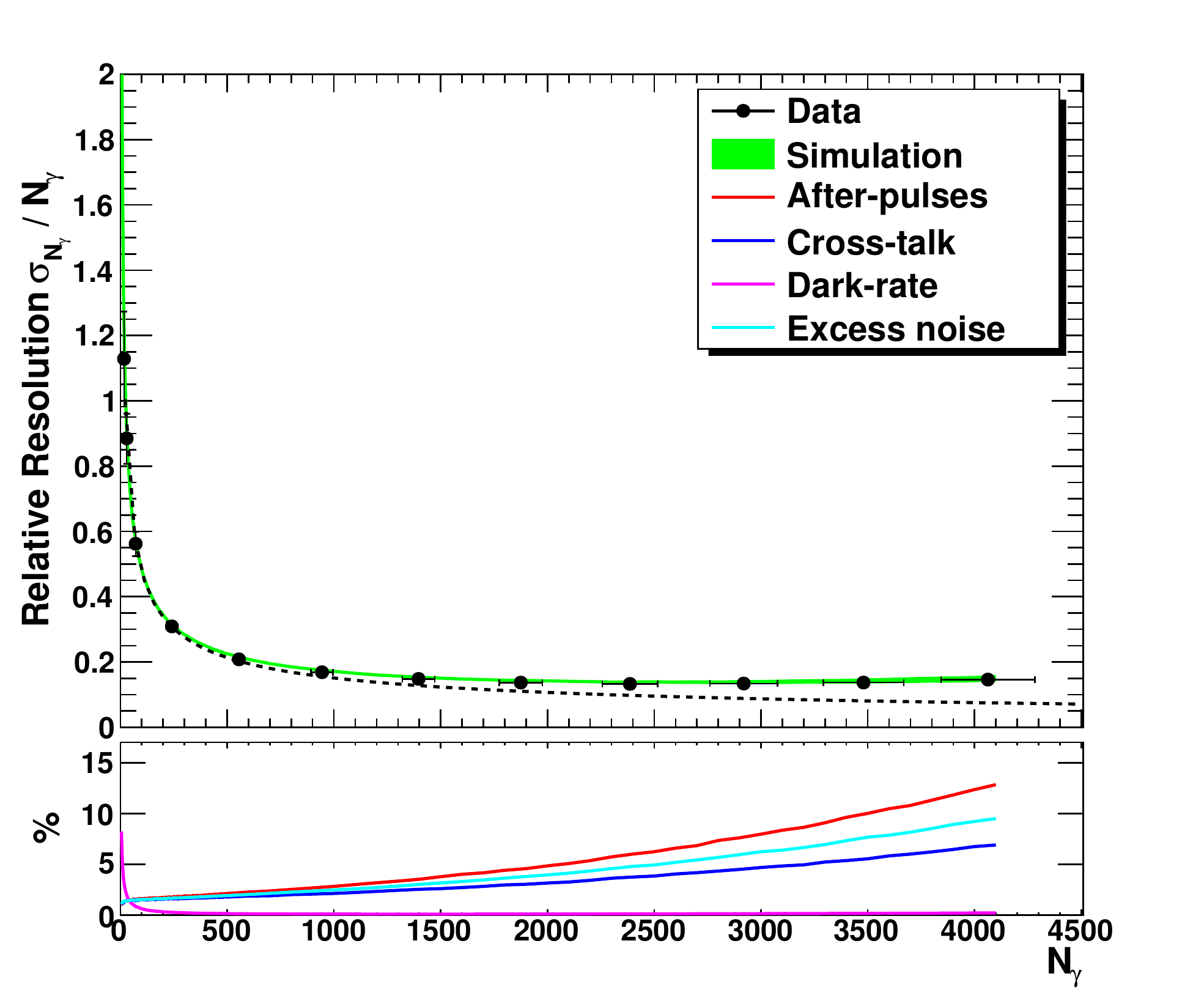}
	\caption{\textbf{Top:} Photon-counting resolution for low noise (0.5\,V over-voltage). The dotted line shows the resolution assuming no saturation effects. \textbf{Bottom:} Contribution from the individual noise sources.}
	\label{fig:resolution0.5}
	
	\includegraphics[width=.75\linewidth]{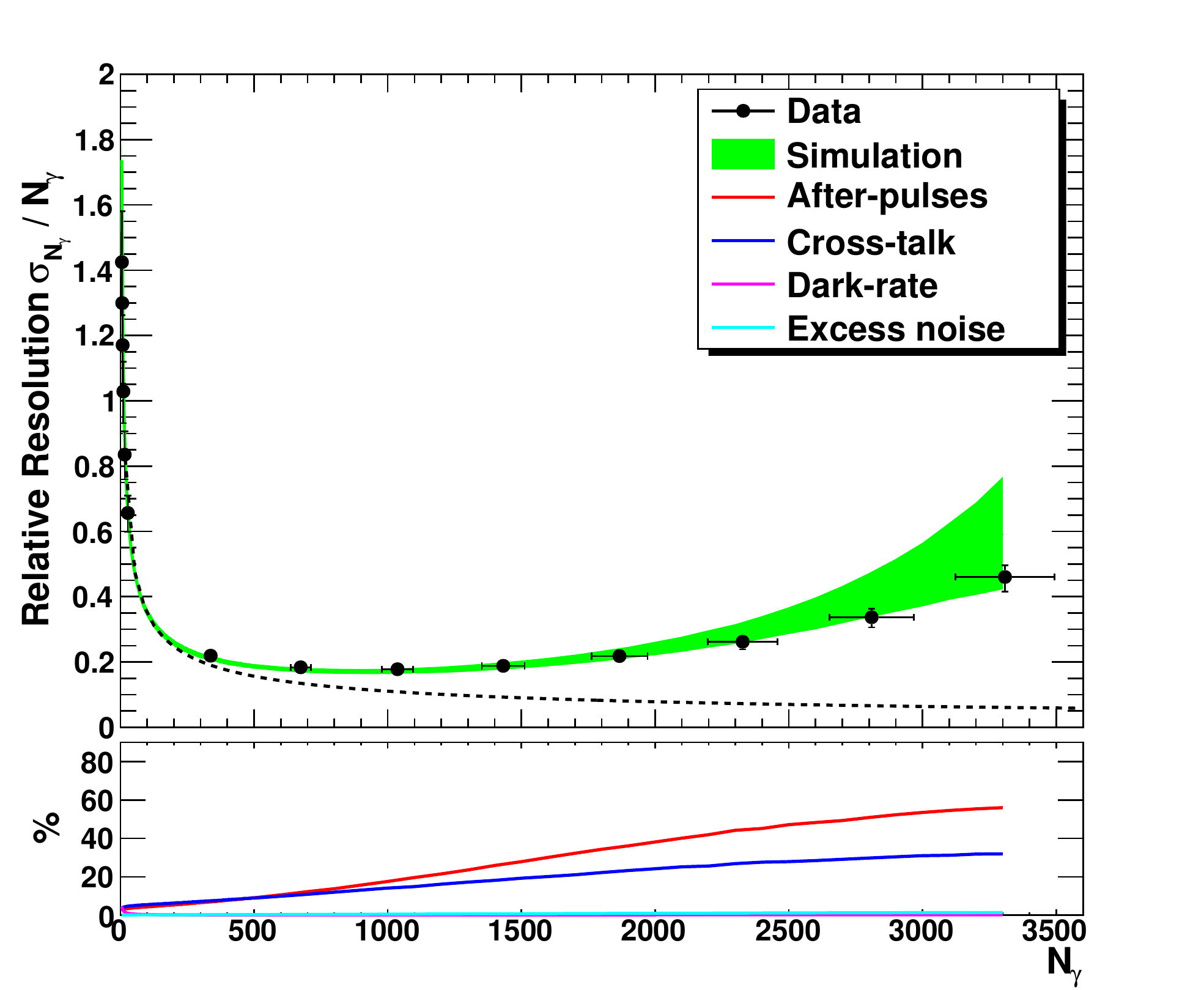}
	\caption{\textbf{Top:} Photon-counting resolution for high noise (1.0\,V over-voltage). The dotted line shows the resolution assuming no saturation effects. \textbf{Bottom:} Contribution from the individual noise sources.}
	\label{fig:resolution1.0}
\end{figure}

Figures~\ref{fig:dynRange0.5},~\ref{fig:rms0.5},~\ref{fig:dynRange1.0},~\ref{fig:rms1.0} show the simulated and measured response curve and the resolution for 0.5\,V and 1.0\,V over-voltage. The contribution from the dominant noise sources determined by the simulation are shown beneath. The error band of the simulated response originates from the measurement uncertainties of the input parameters.
\newline
For 0.5\,V over-voltage, the response curve is fitted using a formula similar to equation~\ref{eqn:responseIdeal}:
\begin{equation}
	N_{av}(N_{\gamma})=N_{pixels} \cdot (1-e^{-\frac{\varepsilon \cdot N_{\gamma}}{N_{pixels}}}) + N_{DR} \mathrm{,}
\end{equation}
where $\varepsilon$ describes the mean number of avalanches, including PDE, cross-talk and after-pulse effects, which are triggered by an incident photon. $N_{DR}$ is the mean number of dark-rate events occurring within the time interval of the integration gate.
For 1.0\,V over-voltage, this formula has to be modified due to the large contribution from cross-talk and after-pulses \cite{MorgunovPrivate}:
\begin{equation}
	N_{av}(N_{\gamma})=N_{pixels} \cdot (1-e^{-\frac{\varepsilon \cdot N_{\gamma}}{N_{pixels}}}) / (1 - \varepsilon_{CN} \cdot e^{-\frac{\varepsilon \cdot N_{\gamma}}{N_{pixels}}}) + N_{DR} \mathrm{,}
\end{equation}
where $\varepsilon_{CN}$ describes the contribution from correlated noise.

For both low and high over-voltage, the simulation reproduces the measured response and resolution within the uncertainties and thus validates the simulation with an accuracy of about $5\%$. For the specific SiPM used (Hamamatsu MPPC S10362-11-100C) the simulation identifies after-pulsing as the dominant source of noise. The contribution from cross-talk to the SiPM response and resolution in the non-linear range is suppressed due to the high pixel occupancy.

The relative photon-counting resolution $\sigma_{N_{\gamma}}/N_{\gamma}$ of a SiPM is determined by the inverted response curve $N_{\gamma}(N_{av})$ and the resolution $\sigma_{N_{av}}$:
\begin{equation}
	\frac{\sigma_{N_{\gamma}}}{N_{\gamma}} = \frac{\partial N_{\gamma}/\partial N_{av} \cdot \sigma_{N_{av}}}{N_{\gamma}} \mathrm{.}
\end{equation}
The relative photon-counting resolution determined via this formula is shown in figure~\ref{fig:resolution0.5} and~\ref{fig:resolution1.0}.
Assuming no saturation effects, the relative resolution can be described by
\begin{equation}
	\frac{\sigma_{N_{\gamma}}}{N_{\gamma}} = \frac{A}{N_{\gamma}} \oplus \frac{B}{\sqrt{N_{\gamma}}} \mathrm{,}
\end{equation}
where $A$ contains the contributions from the dark-rate, and $B$ represents the stochastic term including fluctuations of the number of detected photons due to a limited PDE as well as contributions from cross-talk and after-pulses.
Figures~\ref{fig:resolution0.5} and~\ref{fig:resolution1.0} also show a fit of this function to the measured resolution using only the low intensity data points corresponding to a response below 20 pixels. The fit is then extrapolated to high intensities. It can be seen, that for the tested sensor, this description is valid up to $\approx 40$ firing pixels, which corresponds to $\approx 1000$ photons for 0.5\,V over-voltage and $\approx 300$ photons for 1.0\,V over-voltage.
For higher intensities, the signal saturation contributes significantly to the relative resolution.
In case of $0.5\,V$ over-voltage, the sensor performs well in the whole measurement range up to 4000 incident photons while the best relative resolution of $15\%$ is achieved at $\approx 2500$ photons ($\approx 70$ fired pixels). For $1.0\,V$ over-voltage, a similar value of $20\%$ at $\approx 80$ fired pixels corresponding to $\approx 1000$ incident photons is achieved. For $\gtrsim 100$ firing pixels the relative resolution is degraded due to the strong saturation of the signal.
\newline
The simulation shows, that in a wide range the dominant contribution to the relative resolution is the limited PDE, whereas the effects of cross-talk and after-pulsing are small. Only in the saturation region cross-talk and after-pulsing contribute significantly to the relative photon-counting resolution, due to the large fluctuations in the number of after-pulse and cross-talk events. In addition, the cross-talk also reduces the dynamic range due to the increased pixel occupancy, which further degrades the relative resolution in the non-linear range.

\section{Summary}
In this paper, a simulation framework of the response of Silicon Photomultipliers is presented. The tool offers a generic and detailed modelling of SiPMs and allows easy integration into custom applications. The simulation is validated in the full dynamic range for a 100 pixel Hamamatsu MPPC sensor operated at low and high bias voltage. The data are described by the simulation within approximately $5\%$ uncertainty. The photon-counting resolution of the sensor is measured as a function of the light intensity and the contributions from cross-talk and after-pulses are determined using the simulation.

\acknowledgments

We thank the CALICE collaboration for the good cooperation and inspiring discussions. This work was funded by the BMBF grant no.\ 05HS6VH1 and the International Research Training Group, "Development and Application of Intelligent Detectors" GRK1039/2.

\end{document}